\documentclass[twocolumn,prb,aps,showpacs,superscriptaddress]{revtex4}
\usepackage[hypertex]{hyperref}
\def\rhos{\rho_\star}
\def\half{{1 \over 2}}
\def\rate{{\cal R}}

\def\GL{{\rm GL}}
\def\rcore{r_C}

\def\vel{u}
\def\corr{f}
\def\im{\mathop{\Im{\rm m}}\nolimits}

\def\Lw{L_{\rm w}}
\def\Ad{A_{\rm d}}
\begin{document}
\title{Quantum phase slips in the presence of finite-range disorder}
\author{Sergei Khlebnikov}
\affiliation{Department of Physics, Purdue University, West Lafayette,
Indiana 47907, USA}

\author{Leonid P. Pryadko} 
\affiliation{Department of Physics, University of California,
  Riverside, California  
92521, USA}

\date{July 26, 2005}

\begin{abstract}
To study the effect of disorder on quantum phase slips (QPS) in 
superconducting wires, we consider the plasmon-only model where
disorder can be incorporated into a first-principles instanton calculation.
We consider weak but general finite-range disorder and compute the formfactor
in the QPS rate associated with momentum transfer. We find that 
the system maps onto dissipative quantum mechanics,
with the dissipative coefficient controlled by the wave
(plasmon) impedance $Z$ 
of the wire and with a superconductor-insulator transition at $Z=6.5$ kOhm.
We speculate that the system will remain in this universality class
after resistive effects at the QPS core are taken into account.
\end{abstract}

\maketitle 

The possibility that quantum fluctuations destroy superconductivity in
thin wires has attracted attention of both experimentalists and
theorists for a long time.  Similarly to Little's analysis of thermal
fluctuations,\cite{Little} one concludes that the requisite quantum 
fluctuation
should be sufficiently large, so as to allow the Ginzburg-Landau (GL)
order parameter $\psi_\GL$ to vanish at the core, and the phase of
$\psi_\GL$ to unwind. Such fluctuations are known as
quantum phase slips (QPS).\cite{Giordano}
On the experimental side, there has been a bit of controversy over precisely
how superconductivity disappears in thin wires at low temperatures. 
Some experiments see a sharp superconducting-insulator transition
(SIT),\cite{sit_exp} while others do not.\cite{no_sit} 

Superconductivity takes place when it is difficult to transfer
momentum from the moving condensate.  In one dimension (1D), a QPS
unwinds a large momentum $P \sim \pi n_s$, where $n_s$ is the linear
superconducting electron density, and this momentum has to go
somewhere.  In uniform superfluids (such as a cold Bose gas in a ring
trap), the requirement of momentum transfer constitutes a major
bottleneck for QPS.\cite{Khlebnikov} In superconducting wires, on the
other hand, there are some easily identifiable sinks of momentum. The
most obvious, and as far as we know the only one that has been
considered in the 
literature, is the normal electrons at the QPS core, which in turn
transfer momentum to the disorder potential.  It appears that in the
existing theory of QPS in wires\cite{GZ} this process is assumed to be
100\% effective, so that no trace of momentum conservation is left in
the QPS rate. One should keep in mind, though, that this result is
obtained using instantons of a disorder-averaged theory.

In this work, we analyze effects of suppressed momentum transfer
explicitly, within a simple model where the QPS rate can be found
from a first-principles instanton calculation---by first obtaining 
the rate for a given disorder configuration and then averaging over
disorder. We consider the general case of weak but finite-range disorder.
Our essential simplifying assumption is that electrons
in the core are not effective in transferring momentum to the lattice,
so that the transfer takes place via the gapless plasmon mode.\cite{Kulik&MS}
In this limit (applicability of which is further discussed below), the
rate can be computed within a plasmon-only effective theory [Eq.~(\ref{L_E})].

Under these assumptions we find that the transition point is determined
by the wave (plasmon) impedance of the wire $Z$.  We find that the
system is in the universality class of the dissipative quantum
mechanics\cite{dissipative} (as opposed to the XY universality class
found in Ref.~\onlinecite{GZ}) and identify a SIT at $Z=\pi/2e^2=6.5$
k$\Omega$.  The Ohmic resistivity of the wire at the 
SIT, for weak disorder (the only case considered here), is small (much smaller
than the normal-state resistivity) and
has a non-universal value that depends on both the strength and
correlation length of disorder.

These results apply in the limit when the normal
resistance $R_{\rm core}$ of the QPS core is effectively infinite.
The effect of a finite $R_{\rm core}$ can be understood as follows.
Plasmons produced by a QPS
can be viewed as charge fluctuations in equivalent transmission lines, 
one such line on each side of the QPS core.  A finite $R_{\rm core}$ will 
shunt the charge separation at the QPS core, thereby reducing the plasmon 
emission. This picture of two
transmission lines shunted by $R_{\rm core}$ suggests that the
universality class will remain the same even when dissipation is
caused mostly by a finite $R_{\rm core}$ (which may very well be the
case for existing experimental samples). The SIT will now be 
controlled by the total impedance formed by $Z$ and
$R_{\rm core}$ connected in parallel and thus occur across a straight line
in the $(1/Z,1/R_{\rm core})$ plane.

Our results rely on a certain amount of impedance matching at the
wire's ends.  We assume that plasmons can leave the wire and go into
the leads. This inhibits quantization of plasmon modes and translates,
technically, into the possibility to consider the temperature $T$ and
the wire length $\Lw$ as independent infrared parameters.  In the
limit of short-range disorder, our results can be compared to those
obtained by Luttinger-liquid methods in Ref.~\onlinecite{GS} and
reproduced in the instanton approach in
Ref.~\onlinecite{Kashurnikov&al}.  Because we use the scaling $T\to 0$
with $\Lw$ fixed (as opposed to $T \propto \Lw^{-1}$ used by those
authors), we obtain a different value of the critical coupling. On the
other hand, our value of the critical coupling coincides with that
obtained in the model where an effective resistor is connected to the
wire's ends,\cite{Buchler&al} provided we substitute $Z$ for the
resistance.

We start with the purely bosonic Euclidean lagrangian density
\begin{equation} 
L_E=\psi^\dagger
\partial_\tau\psi+{1\over 2M}|\partial_x\psi|^2+{g\over
  2}|\psi|^4-(\mu+V)|\psi|^2 \; ,
\label{L_E}
\end{equation} essentially the 1D
Gross-Pitaevskii 
model in the presence of the disorder potential $V\equiv V(x)$.
The field $\psi$ is the ``field of Cooper pairs'' describing fluctuations
of superconducting density and phase. Thus, 
$4 \langle \psi^\dagger \psi \rangle m_* / M = n_s$ 
($m_*$ is the effective electron mass) is the 
density of superconducting electrons, while the coupling constant in 
Eq.~(\ref{L_E})
is $g = 4e^2 / C$, where $C$ is the wire capacitance per unit length.
The effective theory
(\ref{L_E}) holds only at spatial scales larger the size of a Cooper pair,
i.e., the GL coherence length $\xi$. Thus, the potential $V(x)$ is 
coarse-grained at the scale $\xi$ in the $x$-direction and at the scale of 
the wire thickness in the transverse directions.

We assume the disorder to be Gaussian with the correlators $\langle
V\rangle=0$, 
$\langle V(x)V(x')\rangle= V_0^2 \corr(x-x')$.  The correlation function
$\corr(x)$ is normalized so that $\corr(0)\equiv 1$, and $V_0$ is the r.m.s.\ 
disorder amplitude.  The disorder correlation length is
$l$, meaning that for $x\agt l$, $\corr(x)\to 0$.

A major role in determining the QPS rate is played by
interactions between QPS at different locations.
These interactions are determined by regions outside the QPS
cores, and can be accounted for in a phase-only model.  We define
$\psi=(\rho+\delta\rho)^{1/2}e^{i\theta}$ and, assuming weak disorder,
expand Eq.~(\ref{L_E}) in powers of the small density fluctuation
$\delta\rho$ around the stationary point with a given phase gradient
$\theta'=\bar\theta'$. This phase gradient takes into account the biasing 
current $I=2e\,\bar\theta'\rho/M$.  At the classical level 
the stationary point is 
characterized by the local minimum $\rho=\rhos$ 
of the effective potential
\begin{equation}
  \label{eq:potential}
  U(\rho)\equiv{g\over 2}\rho^2-\mu \rho-{MI^2\over 8e^2\rho}.
\end{equation}
The minimum exists below the critical current, $I<I_c$, where $3 [M
  I_c^2g^2/(4e^2)]^{1/3}=2\mu$.  

Integrating out the density fluctuations $\delta \rho$, we obtain the
Euclidean lagrangian density for the phase fluctuations $ \theta_1\equiv
\theta-\bar\theta$, describing gapless plasmons\cite{Kulik&MS} propagating 
with speed $c_0=(\rhos g/M)^{1/2}$ on the background moving with
superfluid velocity $\vel\equiv I/(2e\,\rhos)$:
\begin{equation} 
  L_E=i\rhos\dot\theta_1+
  \half D_\tau \theta_1\, \hat K^{-1}\,D_\tau \theta_1
  +  {\rhos\over 2M}
  (\theta'_1+\bar\theta')^2  ,
  \label{eq:exp}
\end{equation}
where $D_\tau\theta_1\equiv \dot\theta_1-i\vel\theta'_1-iV$ is the
covariant time derivative in the moving reference frame and
$\hat{K}\equiv g\,(1-r_s^2 \nabla^2)$ is the differential operator
with the screening length $r_s=(4 M\rhos g)^{-1/2}$. For realistic values of 
the parameters, this screening length
is much smaller than the GL coherence length $\xi$.  Our starting point 
(\ref{L_E}) is already coarse-grained at scale $\xi$; in what follows we set
$\hat{K} = g$.

We can now find the exponential factor in the QPS rate by computing
the action of a suitable classical configuration. We begin with the
case of strictly zero temperature, $T=0$. The leading
effect is due to a single phase slip-antislip pair, or
equivalently a vortex-antivortex pair in the $(x, \tau)$ plane. Away
from the vortex cores
\begin{equation}
  e^ {2i\theta_1}={x-x_0+iv_+(\tau-\tau_0)\over x-x_0-iv_-(\tau -\tau_0)}
  \cdot { x-x_0'-iv_-(\tau-\tau_0')\over
      x-x_0'+iv_+(\tau-\tau_0')},\label{eq:slip-antislip}
\end{equation}
where $v_\mp\equiv c_0\mp \vel$ are the up(down)-stream velocities.
For $u=0$ (i.e., $v_+ = v_-$), this is the configuration familiar 
from the studies of the planar XY model.\cite{BKT}

Integrating the Euclidean lagrangian density (\ref{eq:exp}) with the
configuration (\ref{eq:slip-antislip}) over $x$ and $\tau$, we obtain
the corresponding classical action
\begin{equation}
  \label{eq:action-decomp}
  S_E=S_1+S_2+S_{\rm dis},
\end{equation}
where the combination of uniform linear in $\theta_1$ terms,
\begin{equation}
  \label{eq:act-berry}
  S_1= i P \Delta x_0 - E \Delta \tau_0 ,
\end{equation}
with $\Delta x_0= x_0-x_0'$, $\Delta \tau_0= \tau_0-\tau_0'$, 
accounts for the Berry phase of each QPS [$P = 2\pi \rhos$ is the
momentum released by unwinding the supercurrent] and the released
energy $E=2\pi \rhos \vel=2\pi I/(2e)$,
\begin{equation}
  \label{eq:act-interact}
  S_2={\pi c_0\over g}\ln \biglb[ {(\Delta x_0+iv_+ \Delta \tau_0)(\Delta
  x_0-iv_- \Delta \tau_0)\over \rcore^2}\bigrb]
\end{equation}
is the plasmon-mediated interaction between the phase slips coming
from the terms quadratic in $\theta_1$, 
and 
\begin{equation}
  \label{eq:act-disorder}
  {S_{\rm dis}}= -{2\pi i\over 
  g(1-\vel^2/c_0^2)} \int_{x_0'}^{x_0}dx\, V(x)
\end{equation}
is the effect of the disorder. In Eq.~(\ref{eq:act-interact}), we have used
the QPS core size $\rcore$ (which is not determined by the present theory)
as the short-distance cutoff.

These expressions illustrate the effect of the superfluid velocity
$\vel$ on the QPS action and can be useful, for instance, in weakly
non-ideal Bose gases, where $\vel$ may in principle approach $c_0$.
Thus, for example, factor
$(1-\vel^2/c_0^2)^{-1}=(v_+^{-1}+v_-^{-1})\,c_0/2$ in
Eq.~(\ref{eq:act-disorder}) arises because the up(down)-stream
plasmons spend longer (shorter) time at the place with the given
density. In thin superconductors, however, we typically have $\vel \ll v_F
\alt c_0$.  In the following we neglect $\vel/c_0\ll1$, to get
\begin{equation}
\label{simpl}
  S_E = S_1 + {2\pi c_0\over g}\ln \frac{\Delta r_0}{\rcore} 
  - {2\pi i\over g} \int_{x_0'}^{x_0}dx\, V(x) ,
\end{equation}
where $\Delta r_0=(\Delta x_0^2+c_0^2\Delta \tau_0^2)^{1/2}$.  We observe
that the last term here can be interpreted as the modification of
$S_1$ due to the local correction to the density
$\delta\rhos(x)=-V(x)/g$ caused by disorder.  We will
assume that $V(x)$ incorporates all mechanisms leading to linear
density inhomogeneity: non-uniform wire cross-section, magnetic
impurities, etc.  The weak-disorder approximation is applicable for 
$|\delta \rhos|\ll\rhos$, which gives the dimensionless measure of the
disorder strength, 
\begin{equation}
\alpha \equiv V_0/M c_0^2 = V_0 / g \rhos \; .
\label{alpha}
\end{equation}

For a given disorder configuration, the QPS rate (per unit length) can
be found as the imaginary part of the partition sum of the pair,
\begin{equation}
  \label{eq:qps-rate}
  \rate={C_1 \,c_0^4\over g^2\rcore^4}\,R,\;\, R\equiv \im \int
  {dx_0\,dx_0'\over \Lw}\,\int d\Delta \tau_0 \,e^{-S_E},
\end{equation}
where the dimensionful prefactor with the coefficient $C_1\sim1$
incorporates the fluctuation determinant and the Jacobian of
transformation to the collective coordinates.  The integration over
$x_0$, $x_0'$ is extended over the length $\Lw$ of the wire.  For
large enough $\Lw$ the rate~(\ref{eq:qps-rate}) is self-averaging with
respect to disorder.  In this case, the effective action is obtained
by disorder averaging,
\begin{equation}
\label{S_eff}
  S_{\rm eff} = S_1 + {2\pi c_0\over g}\ln \frac{\Delta r_0}{\rcore}
  + \half P^2 \alpha^2 \Delta(x_0-x_0') ,
\end{equation}
where the function
\begin{equation}
  \Delta(x) 
  = 2\!\int_0^{x}\!\!  dx'\, (x-x')\,\corr(x')
  \label{eq:delta}
\end{equation}
is proportional to the average square of the disorder-induced phase between
the slip and antislip.  [The disorder correlator $\corr(x)$ is defined after 
Eq.~(\ref{L_E}).] For finite-range disorder, $\Delta(x)$
has a diffusive form at large 
distances, $|x| \gg l$,
\begin{equation}
  \label{eq:lengths}
  \Delta(x)=
  2l_1 |x|,\;\;\;\,
  l_1\equiv \int_0^\infty \!\!\!\!dx\, \corr(x).
\end{equation}
For single-scale disorder, $l_1\sim l$.

According to Eqs.~(\ref{S_eff}), (\ref{eq:lengths}), disorder
binds together the $x$ coordinates of the phase slip and anti-slip within a
pair. For the case of short-range disorder, this effect was noted previously
in Ref.~\onlinecite{Kashurnikov&al} and was taken into account by using
a sharp $\delta$-function, $\delta(x_0-x_0')$, in the partition sum.
We see, however, that in general
there is a finite characteristic separation $\Delta x_0$
within the pair. This 
separation depends both on the range and on the strength of the disorder 
potential. As the strength of disorder increases, and the pairs become more
tightly bound in the $x$ direction, the wire
effectively becomes more and more like a dissipative
quantum-mechanical system with a logarithmic interaction, $\ln |\tau_0
-\tau'_0|$, 
between the instantons---a spatially extended version of the familiar 
resistively-shunted Josephson junction.\cite{dissipative}

We also note that the coefficient in front of the logarithm in
Eq.~(\ref{S_eff}) is proportional to the inverse impedance of the wire
viewed as two transmission lines attached to the phase-slip region.
This is consistent with the mapping onto dissipative quantum
mechanics, since plasmons are the only source of dissipation in our
present model.

Turning to the time integration in Eq.~(\ref{eq:qps-rate}), we observe that
while the integral is formally
divergent, the corresponding imaginary part is finite and can be found by
an analytic continuation. We obtain
\begin{equation}
  \label{eq:qps-rate-integrated}
  R={\pi^{3/2}\rcore\over c_0}\!
  \int\!\! {dx}\, e^{iP x-\half{\alpha^2}P^2\Delta(
  x)} \Bigl({E\rcore^2\over 2 c_0
  |x|}\Bigr)^{\nu} {J_\nu(E|x|/c_0)\over \Gamma(\nu+1/2)},
\end{equation}
where $J_\nu(z)$ is the Bessel function and the dimensionless index
$\nu$ is inversely proportional to the interaction constant,
$\nu+1/2\equiv \pi c_0/g$ ($=K^{-1}$ in notations of
Ref.~\onlinecite{GS}). This explicit expression for the QPS rate  
(at $T=0$) in 
the plasmon-only theory in the presence of weak but finite-ranged
disorder is the main result of this work. Although 
Eq.~(\ref{eq:qps-rate-integrated}) was obtained
assuming that disorder induces only weak fluctuations of the density
(as expressed by the weak-disorder condition $\alpha \ll 1$), 
it is non-perturbative with respect to the phase
fluctuations and accounts for multiple scattering to all orders.

In the
absence of disorder [$\alpha = 0$] the integral 
in Eq.~(\ref{eq:qps-rate-integrated}) is zero for any
$E < c_0 P$, consistent with the Galilean invariance of the $T=0$ 
state.\cite{Khlebnikov}
The opposite case is when the convergence of the integral
(\ref{eq:qps-rate-integrated}) is dominated by disorder. In this case, 
the Bessel function can be replaced by the first term of 
its small-argument expansion,
and the QPS rate becomes
\begin{equation}
  \label{eq:qps-rate-small}
  \!\!\rate={C_2 c_0^3\over g^2 \rcore^3}\Bigl({E\rcore\over 2
  c_0}\Bigr)^{2\nu}\!\! l_1\Ad,\; \Ad\equiv\!
  \int\! {dx\over l_1}\, {\displaystyle e^{iP x-\half
  P^2\alpha^2\Delta(x)}}\!\! 
\end{equation}
with $C_2=\pi^{3/2}\,C_1/[\Gamma(\nu+1)\Gamma(\nu+1/2)]$. 
This limit corresponds to the entire momentum $P$ being absorbed by the
disorder, with no momentum carried away by plasmons. For weak
disorder, where Eq.~(\ref{eq:qps-rate-small}) is applicable, the
dimensionless formfactor $\Ad$ is small.

At nonzero temperature, the simplest case is $T \gg gE/\pi c_0$. Then, 
the energy
released by unwinding the supercurrent is insignificant, and instead of 
a single pair (\ref{eq:slip-antislip}) we can use a periodic chain of 
such pairs with period $\beta = 1/T$. 
Properties of this chain are described in Ref.~\onlinecite{Khlebnikov}. 
An especially simple result applies in the disorder-dominated regime,
when the spatial separation in each pair is small,
$\Delta x_0 \ll c_0 / \pi T$: for an estimate, 
it is sufficient to make the replacement $E\to \pi c_0 T / g$ 
in Eq.~(\ref{eq:qps-rate-small}). Then, at $\nu \sim 1$, the rate 
of thermally-assisted QPS per unit length is
\begin{equation}
\rate_T \sim {c_0 \over \rcore^3}\Bigl({T\rcore\over g}
  \Bigr)^{2\nu}\!\! l_1\Ad \; .
\label{rate_T}
\end{equation} 
The voltage across the wire is
\begin{equation}
V = \frac{2\pi}{e} \rate_T \Lw \sinh \frac{\pi I}{2e T} \; .
\label{volt}
\end{equation}
We see that resistance becomes $T$-independent at $\nu= 1/2$, 
which defines  the SIT 
point. The dimensionless measure of QPS pair density near $\nu= 1/2$ is
\begin{equation}
\rate_T \Delta x_0 \Delta \tau_0 \sim \frac{l_1 \Delta x_0}{\rcore^2} 
\Bigl({T \rcore\over g}\Bigr)^{2\nu-1} \Ad \; ,
\label{dens}
\end{equation}
where $\Delta \tau_0 = 1/2T$ is the characteristic size of a pair
in the $\tau$ direction. Thus, the density is small in the infrared at any
$\nu > 1/2$, and for sufficiently small $\Ad$ even at $\nu=1/2$. At 
$\nu < 1/2$, pairs proliferate, resulting in the insulating behavior.
Writing the impedance of the wire as $Z = 2(L/C)^{1/2}$, where
$L= m_* / e^2 n_s$ is the ``kinetic'' inductance per unit length, and
the factor of $2$  
corresponds to two transmission lines (one on each side of the phase slip), 
we see that $\nu = 1/2$
is equivalent to $Z = \pi/ 2e^2=6.5$ k$\Omega$. 

We now turn to discussion of the formfactor $\Ad$. The important
parameter here is $\alpha^2 P l$. We begin with the case $\alpha^2 P l \ll 1$.
If the stronger condition $\alpha P l \ll 1$ is also satisfied, we can 
(upon integrating by parts twice) expand the integrand in 
(\ref{eq:qps-rate-small}) to the linear order in $\alpha^2$. 
This corresponds to the entire momentum
$P$ being absorbed in a single scattering event. Then,
$\Ad= (\alpha^2 /l_1) \int e^{iPx} \corr(x) dx$. Thus, for $P l \ll 1$
the formfactor is universal, $\Ad= 2\alpha^2$, while for $l\agt P^{-1}$ it
depends on details of the correlations.  For $Pl\gg1$, it is determined by the
ordinate $l_0$ of the singularity of $\corr(x)$ closest to the real axis
and scales exponentially, $\Ad\propto e^{- Pl_0}$. If $\alpha P l \agt 1$
(but still $\alpha^2 P l \ll 1$), effects of multiple scattering result in
a correction $\sim \alpha^2 P^2 l^2$ in the exponent.  

In the opposite limit, $\alpha^2 Pl\gg1$, it is the disorder that
determines convergence of the integral (\ref{eq:qps-rate-small}). The integral
converges at $x\sim (\alpha^2 P)^{-1}\ll l$, where we can replace 
$\Delta(x)$ by its short-distance form, $\Delta(x)=x^2$. Then, the integral 
becomes Gaussian and gives
$\Ad=(2\pi)^{1/2}( \alpha Pl_1)^{-1} e^{-1/(2\alpha^2)}$.
In this case the process of momentum transfer to disorder is clearly 
a result of a large number of scattering events.

The nature of the crossovers between different scattering regimes as
the disorder correlation length $l$ is increased can be understood in a model
with correlation function $\corr(x)=(1+x^2/l^2)^{-3/2}$.  Integration
in Eq.~(\ref{eq:delta}) gives  
$\Delta(x)=2l^2[(1+x^2/l^2)^{1/2}-1]$, and the coordinate
integration in Eq.~(\ref{eq:qps-rate-integrated}) can be performed
explicitly,
$$
  \Ad= {2 \alpha^2 Pl\over [1+  (\alpha^2
  Pl)^2]^{1/2}}\,e^{\alpha^2 P^2 l^2} K_1\biglb(Pl[1+  (\alpha^2
  Pl)^2]^{1/2}\bigrb). 
$$
For $Pl\ll 1$, we can use $K_1(z)=z^{-1}+\mathcal{O}(z\ln z)$
which gives the short-range limit (with some logarithmic corrections
due to the power-law tail of the correlation function), while for
$Pl\gg1$, the large-argument asymptote $K_1(z)\approx
(\pi/2z)^{1/2}e^{-z}$ can be used to restore the other discussed
scattering regimes.  

Our discussion of $\Ad$ has so far assumed that the scattering is characterized
by a single distance scale, $l$, which may not necessarily be the
case.  For example, a model correlation function $\corr(x)=e^{-|x|/l}$ has
a singularity (derivative discontinuity) at the real axis, which gives
distance $l_0=0$, and the exact integration in Eq. (\ref{eq:qps-rate-small})
shows that $\Ad$ is independent of $P$ for the
entire range $\alpha P l\ll1$.  Conversely, for a correlation function
with the power-law long-distance tail, $\corr(x)\propto x^{-m}$, $0<m<1$,
the distance $l_1$ is infinite, and instead of 
Eq.~(\ref{eq:lengths}) one obtains superdiffusive form
$$\Delta(x)= {l_*^2\over 2-m}\Bigl({x\over l_*}\Bigr)^{2-m},\quad  x\gg
l_*, 
$$
where $l_*$ is a convenient distance scale.  Then, perturbation
theory breaks down already for arbitrarily weak disorder; we get
$\ln( \Ad)\propto -[\alpha^2 (Pl_*)^m]^{-m/(1-m)}$.

In conclusion, the plasmon-only theory provides a useful laboratory 
for studying
the effect of disorder on QPS from first principles. We have computed the
QPS rate for general finite-range weak disorder; as seen from
Eq.~(\ref{eq:qps-rate-integrated}), disorder drives 
the system into the universality class of dissipative quantum mechanics, with 
the dissipative coefficient determined by the wave impedance $Z$
of the wire. From an extension of this result to finite temperatures,
Eq. (\ref{rate_T}), 
we have found a superconductor-insulator transition at $Z = \pi/2e^2$.
We see no reason why the universality class should change when resistive 
effects at the QPS core are taken into account. Rather, we expect, that it
will remain the same, but the dissipative coefficient will now be determined
by the total impedance, including both plasmon and resistive effects.

We thank A.\ Bezryadin and S.\ Chakravarty for useful discussions, and
the Aspen Center for Physics, where this work was started, for
hospitality.  S.K.\ was supported in part by the U.S.\ Department of
Energy through Grant DE-FG02-91ER40681 (Task B).

\end{document}